\documentclass{article}%
\usepackage{amsmath}%
\usepackage{amsfonts}%
\usepackage{amssymb}%
\usepackage{graphicx}

\begin{document}

\title{On the differential equation $y^{''}-\frac{b^{'}(x)}{2b(x)}y^{'}+\lambda b(x)y=0$}
\author{Christian Rakotonirina
\\Institut Sup\'{e}rieur de Technologie d'Antananarivo, IST-T, Madagascar\\
e-mail:rakotopierre@refer.mg}

\maketitle
\begin{abstract}
The Hamilton-Jacobi method which can be used for solving this equation has been presented. The solution of the equation suggests that there exist some second order linear ordinary differential equations whose resolution can be done by means of characteristic equation.   \end{abstract}
\maketitle
\section{Introduction}
As the title show, in this paper we deal with second order linear ordinary differential equation (ODE). There is not general method for resolving all second order linear ODE's. The method depend on the form of the equation. Moreover, there exist second order linear ODE's which can not be solved by the existing methods.\\
The Hamilton-Jacobi (H-J) method is a method very useful for solving differential equation of the classical mechanics. Knowing that the differential equation of the classical mechanics are all at most second order differential equations, it is of course to ask the question: what are the second order linear ODE's can be solved by the H-J method?\\
For giving response to this question, we tried to apply the H-J method to solve the  most general form of second order linear ODE. During the resolution we moved apart the cases which are difficult to solve. Then, we have at last that the differential equation
\begin{equation}\label{eqn1}
y^{''}-\frac{b^{'}(x)}{2b(x)}y^{'}+\lambda b(x)y=0
\end{equation}
 can be solved by the H-J method. For the H-J method we only consider the case of only one degree of freedom. In the second section we talk about the H-J method and its application to equation (\ref{eqn1}). In the last section we give a set of second order ODE's whose resolution use characteristic equation.
\section{Resolution with H-J method}
\subsection {H-J method}
The H-J method \cite{LandauLifchitz81} is used to solve the equation of motion, which is the Euler-Lagrange equation
\begin{equation}\label{eqn2}
 \frac{d}{dx}(\frac{\partial L}{\partial y^{'}})-\frac{\partial L}{\partial y}=0
 \end{equation}
 associated to the lagrangian $L=L(x,y,y^{'})$, which is a class $\mathcal{C}^3$ function of his three arguments. This equation can be written under the form \cite{Berest93}
\begin{equation}\label{eqn3}
y^{''}\frac{\partial^{2} L}{\partial y^{'2}}+y^{'}\frac{\partial^{2} L}{\partial y^{'}\partial y}+\frac{\partial^{2} L}{\partial y^{'}\partial x}-\frac{\partial L}{\partial y}=0
 \end{equation}

We resume here step by step this method.

\begin{description}
\item[step 1] Constructing the hamiltonian $H=H(x,y,\frac{\partial L}{\partial y^{'}})=y^{'}\frac{\partial L}{\partial y^{'}}- L$
\item[step 2] Solving with respect to $S=S(x,y,\alpha)$ the H-J equation $\frac{\partial S}{\partial x}+H(x,y,\frac{\partial S}{\partial y})=0$, where $\frac{\partial S}{\partial y}$ replaces $\frac{\partial L}{\partial y^{'}}$.
\item[step 3 ] Finally, the resolution with respect to $y$ of the equation $\frac{\partial S}{\partial \alpha}=\beta$ gives the general solution.
\end{description}

\subsection{Search for lagrangian}
These steps show us that knowing the lagrangian we can try to apply the H-J method for solving the equation of motion, that is  the Euler-Lagrange equation. So, we will be able to apply the H-J method to solve an equation of the form
\begin{equation}
y^{''}=f(x,y,y^{'})\label{eqn4}
\end{equation}
 if we can have a function $L=L(x,y,y^{'})$, class $\mathcal{C}^3$ function of his three arguments, such that the equation  is equivalent to the Euler-Lagrange equation associated to the function $L$, as lagrangian. \\
 For searching for such function $L$ \cite{Berest93}, replace $y^{''}$ by $f(x,y,y^{'})$ in (\ref{eqn3}), we have $f\frac{\partial^{2} L}{\partial y^{'2}}+y^{'}\frac{\partial^{2} L}{\partial y^{'}\partial y}+\frac{\partial^{2} L}{\partial y^{'}\partial x}-\frac{\partial L}{\partial y}=0$.\\

 Take the derivative of this identity with respect to $y^{'}$, and in using again the relation (\ref{eqn4}), we have that the necessary condition for
to be of the form of the Euler-Lagrange  equation is

\begin{equation}
-\frac{\partial f}{\partial y^{'}}=\frac{d}{dx}Log\left(\frac{\partial^{2} L}{\partial y^{'2}}\right)
\end{equation}
\subsection{Resolution of the equation (\ref{eqn1})}
For (\ref{eqn1}) we have as a lagrangian $L=\frac{1}{2}\left(-\lambda b(x)y^{2}+y^{'2}\right)\sqrt{\left|b(x)\right|}$.
Now, we are in a position to use the H-J method for solving the equation (\ref{eqn1}). Then, by applying the H-J method, we will have that the equations
\begin{equation}
y^{''}-\frac{b^{'}(x)}{2b(x)}y^{'}+\omega^{2}b(x)y=0
\end{equation}
and
\begin{equation}
y^{''}-\frac{b^{'}(x)}{2b(x)}y^{'}-\omega^{2}b(x)y=0
\end{equation}
with $\omega$ real constant, have respectively as general solutions
\begin{equation}
y=Asin\left[\omega t(x)+\varphi\right]
\end{equation}
$A$, $\varphi$, real arbitrary constants,
and
\begin{equation}
y=Ae^{\omega t(x)}+Be^{-\omega t(x)}
\end{equation}
$A$, $B$, real arbitrary constants, and
where $t(x)=\int\sqrt{\left|b(x)\right|}dx$.
\section{Characteristic equation}
They are a generalisation of the equations $y^{''}+\omega^{2}y=0$ and $y^{''}-\omega^{2}y=0$ whose general solutions are respectively $y=Asin\left[\omega x+\varphi\right]$ and $y=Ae^{\omega x}+Be^{-\omega x}$.\\
For the resolution of these equations one write at first the characteristic equation which are respectively $\lambda^{2}+\omega^{2}=0$ and $\lambda^{2}-\omega^{2}=0$. This method can be applied for all homogenous linear ODE's with constant coefficients. That makes us to search for the second order linear ODE whose fundamental solutions are $e^{\alpha t(x)}$ and $e^{\beta t(x)}$, where $\alpha$, $\beta$ are constant complex numbers and $t(x)=\int\sqrt{b(x)}dx$, with $b(x)$ a positive function. Such equation is
\begin{equation}
y^{''}-\left[\left(\alpha+\beta\right)\sqrt{b(x)}+\frac{b^{'}(x)}{2b(x)}\right]y^{'}+\alpha\beta b(x)y=0
\end{equation}
So, for the equation
\begin{equation}
y^{''}+\left[\mu\sqrt{\left|b(x)\right|}+\frac{b^{'}(x)}{2b(x)}\right]y^{'}+\nu b(x)y=0
\end{equation}
the characteristic equation is $\lambda^{2}+\mu\lambda+\nu=0$ and the general solution is $y=y=Ae^{\lambda_{1} t(x)}+Be^{\lambda_{2} t(x)}$, with $\lambda_{1}$, $\lambda_{2}$ the roots of the characteristic equation. and $t(x)=\int\sqrt{b(x)}dx$.
\subsection*{Acknowledgements} The author would like to thank
Rabearivelo Patrice for discussion and P. Berest for his kindness of sending me, in 1993, one of his articles and book \cite {Berest93}, when in Madagascar internet did not exist yet.

\end{document}